\documentclass{elsart}
\usepackage[dvips]{graphicx}
\journal{Physics Letters A}
\begin{document}

\begin{frontmatter}

\title{Exact Localized Solutions of 
Quintic Discrete Nonlinear Schr\"odinger Equation}

\author[a,b]{Ken-ichi Maruno},
\ead{maruno@riam.kyushu-u.ac.jp}
\author[c]{Yasuhiro Ohta} and
\author[d]{Nalini Joshi}

\address[a]{Research Institute for Applied Mechanics, 
Kyushu University,\\ Kasuga, Fukuoka, 816-8580, Japan}

\address[b]{
Optical Sciences Centre, Research School of Physical
 Sciences and Engineering,\\ 
The Australian National University, Canberra ACT 0200, 
Australia
}

\address[c]{Department of Applied Mathematics,Faculty of Engineering,\\
Hiroshima University, 1-4-1 Kagamiyama, Higashi-Hiroshima 739-8527, Japan}

\address[d]{
School of Mathematics and Statistics,\\
University of Sydney, NSW 2006, Australia
}

\date{\today}

%\maketitle

\begin{abstract}
We study a new quintic discrete nonlinear Schr\"odinger (QDNLS)
equation which reduces naturally 
to an interesting symmetric difference 
equation of the form $\phi_{n+1}+\phi_{n-1}=F(\phi_n)$.  
Integrability of the symmetric mapping is checked by 
singularity confinement criteria and growth properties. 
Some new exact localized solutions for integrable cases 
are presented for certain sets of parameters. 
Although these exact localized solutions represent 
only a small subset of the large
variety of possible solutions admitted by the QDNLS equation, 
those solutions presented here are the first example of
exact localized solutions of the QDNLS equation. 
We also find chaotic behavior for certain parameters of 
nonintegrable case. 
\end{abstract}
%\vspace*{2.0truecm}

\end{frontmatter}

\section{Introduction}
Discrete solitons in nonlinear lattices 
have been the focus of
considerable attention in diverse branches of science
\cite{FWreport,HTreport,Bishop}. 
Discrete solitons are possible in several physical settings, 
such as biological systems \cite{Scott}, atomic chains \cite{Scott2,Takeno}, 
solid state physics \cite{SSH}, electrical lattices \cite{MBR} and 
Bose-Einstein condensates \cite{TS2001}.
Recently, the existence of discrete solitons in photonic structures 
(in arrays of coupled nonlinear optical waveguides 
\cite{Lederer,CJ88,ESMBA,Kivshar,Bang,Musslimani,Sukhorukov1} 
and in a nonlinear photonic
crystal structure \cite{Christodoulides2002}) was 
announced and has attracted considerable attention
in the scientific community.
Photonic crystals, which are artificial microstructures having photonic
bandgaps, can be used to precisely control propagation of optical pulses
and beams. They are very useful for optical components such as 
waveguides, couplers, cavities and optical computers. 
It is possible to make discrete waveguides using photonic crystals. 
In this situation, ``discrete solitons'' appear naturally and have interesting
properties. Many scientists believe that 
the discrete solitons can have an important role in this technology.

%In these systems, discrete localized states are possible 
%%by balancing the effect
%of discrete diffraction (arising from the coupling between neighboring
%waveguides) with that of material nonlinearity. 

Here, we consider the physical system described by 
the discrete nonlinear Schr\"odinger (DNLS) equation. 
Most relevant for realistic applications (particularly to
nonlinear optics) is the DNLS equation with a cubic (Kerr)
nonlinearity:
\begin{equation}
 i\frac{d\psi_n}{dt}+\alpha (\psi_{n+1}-2\psi_n+\psi_{n-1})
+\beta |\psi_n|^2 \psi_n=0\,,\label{dnls}
\end{equation}
where $\psi_n$ are complex variables defined for all integer values of
the site index $n$. 
The DNLS equation was used by Christodoulides and
Joseph \cite{CJ88} to model the propagation of
discrete self-trapped beams in an array of weakly-coupled nonlinear
optical waveguides. In such an array, when low intensity light 
is injected into one, it will couple to more and more
waveguides as it propagates, thereby broadening its spatial
distribution (diffraction). 
High intensity light changes the refractive index of the
input waveguides through the Kerr effect and decouples them from the
rest of the array. Certain light distributions
propagate while retaining a fixed spatial profile among a limited number
of waveguides. These are
discrete spatial solitons. Experimental results for optical waveguide arrays,
confirming the validity of the model, have been reported by
Eisenberg et al. \cite{ESMBA}.
%%It is also often used as an
%%envelope equation modeling the local denaturation of the DNA double
%%strand, and as an equation modeling Bose-Einstein condensation. 

There are many works about the DNLS equation and some its modifications. 
It is well-known that the standard DNLS equation is not completely
integrable \cite{RT}. 
The integrable discrete nonlinear Schr\"odinger equation (Ablowitz-Ladik
(AL) system), 
\begin{eqnarray}
i\frac{d\psi_n}{dt}+\alpha (\psi_{n+1}-2\psi_n+\psi_{n-1})
+\gamma|\psi_n|^2(\psi_{n+1}+\psi_{n-1})=0\,,
\end{eqnarray} 
was proposed by using inverse scattering method \cite{AL}.
The AL system has $N$-soliton solutions and a rich mathematical structure. 
However, it is not physically realistic, because 
it does not contain the Kerr-nonlinearity term. 
Thus the deformed DNLS equation
\begin{eqnarray}
i\frac{d\psi_n}{dt}+\alpha (\psi_{n+1}-2\psi_n+\psi_{n-1})
+\beta |\psi_n|^2 \psi_n
+F(\psi_{n+1},\psi_n,\psi_{n-1})=0,\label{gpdnls}
\end{eqnarray}
where $F(x,y,z)$ is a polynomial function of $x,y,z$, 
was proposed in several works.
For example, 
\begin{eqnarray}
%%&& i\frac{d\psi_n}{dt}+\alpha (\psi_{n+1}-2\psi_n+\psi_{n-1})
%%+\beta |\psi_n|^2 \psi_n\nonumber \\
%%&&\quad \quad +
F(\psi_{n+1},\psi_n,\psi_{n-1})\equiv
\gamma|\psi_n|^2(\psi_{n+1}+\psi_{n-1})\label{pdnls}
\end{eqnarray}
was considered by several authors \cite{Salerno,KS94,CBG94,CBG96,Henning}. 
Equation (\ref{gpdnls}) with Eq.(\ref{pdnls}) 
can be viewed as a deformation of the DNLS equation (\ref{dnls}).
%%%\begin{equation}
%%% i\frac{d\psi_n}{dt}+\alpha (\psi_{n+1}-2\psi_n+\psi_{n-1})
%%%+\beta |\psi_n|^2 \psi_n=0.\label{sdnls}
%%%\end{equation}
Indeed, by using the parameterization $\gamma=(\beta'-\beta)/2$, one
can see that, for $\beta=\beta'$, 
Eq.(\ref{gpdnls}) with Eq.(\ref{pdnls}) reduces to the
standard DNLS equation (\ref{dnls}), 
while for $\beta \to 0$ it gives the AL system. 
Kivshar and Salerno explained the physical meaning of Eq.(\ref{gpdnls})
with Eq.(\ref{pdnls}) \cite{KS94}. 
They proposed that an application of Eq.(\ref{gpdnls})
with Eq.(\ref{pdnls}) may be
found in nonlinear optics in, that the DNLS equation describes 
interactions of partial TE (Transverse electric) 
modes in array of
(focusing or defocusing) waveguides. Thus, in spite of the fact that the
AL system itself is not physical, it can appear as a particular
case of a more general and physically better justified discrete model
Eq.(\ref{gpdnls}) with (\ref{pdnls}). 

The existence of localized solutions 
is well known for spatially discrete systems 
(including quintic and higher order nonlinearities) 
\cite{MacKay,Flach,Malomed,Weinstein}. 
Moreover the stability of these solutions 
at weak coupling is established
as well, 
independent of the integrability of 
the underlying equations of motion. 

However, nobody has found exact soliton-like solutions of 
Eq.(\ref{gpdnls})-type discrete systems. 
Concrete forms of localized solutions are, 
definitely, useful for analyzing real physics. 
Moreover, from physical point of view, 
investigating what kind of Eq.(\ref{gpdnls}) has 
exact localized solutions and what kind of solution 
exits are important. 
Thus the problems we want to address here are: 
Does Eq.(\ref{gpdnls}) with Eq.(\ref{pdnls}) have
soliton-like solutions and what kind of deformed DNLS 
permits soliton-like solutions?

Here, we propose a new model of the propagation of discrete self-trapped
beams in an array of weakly-coupled nonlinear optical waveguides, 
the quintic discrete nonlinear Schr\"odinger (QDNLS) equation  
\begin{eqnarray}
&&i\frac{d\psi_n}{dt}+\alpha (\psi_{n+1}-2\psi_n+\psi_{n-1})
+\beta |\psi_n|^2 \psi_n\nonumber\\
&&\quad \quad +\gamma|\psi_n|^2(\psi_{n+1}+\psi_{n-1})
+\delta|\psi_n|^4(\psi_{n+1}+\psi_{n-1})=0.\label{newdnls}
\end{eqnarray}
In physical problems, the quintic nonlinearity can be of equal or even
higher importance to the cubic one \cite{Moores93} as it is responsible
for stability of localized solutions. 
This equation (in normalized coefficients 
$\alpha=1,\gamma=-4\delta/\beta$) 
can be derived from the Hamiltonian
\begin{equation}
 H=-\sum_n(\psi_n \psi_{n+1}^*+\psi_n^*\psi_{n+1})-
\frac{\beta}{2\delta}\sum_n \ln 
(1+\gamma |\psi_n|^2+\delta |\psi_n|^4),
\end{equation}
with the deformed Poisson brackets
\begin{equation}
\{\psi_n,\psi_m^*\}
=i(1+\gamma |\psi_n|^2+\delta |\psi_n|^4)\delta_{nm}\,,\quad
\{\psi_n,\psi_m\}=\{\psi_n^*,\psi_m^*\}=0\,.
\end{equation}
In general,
\begin{equation}
 \{B,C\}=i\sum_n\left(
\frac{\partial B}{\partial \psi_n}
\frac{\partial C}{\partial \psi_n^*}
-
\frac{\partial C}{\partial \psi_n}
\frac{\partial B}{\partial \psi_n^*}
\right)(1+\gamma |\psi_n|^2+\delta |\psi_n|^4)\,.
\end{equation}
The equation of motion is 
\begin{equation}
\dot{\psi}_n=\{H,\psi_n\}\,. 
\end{equation}

Surprisingly, the QDNLS equation has exact soliton solutions 
(bright and dark) and Jacobi elliptic function solutions, as we shall show
here. 

In this letter, 
we discuss the integrability of mappings which are reduced from the
QDNLS equation and exact localized solutions of the QDNLS equation. 
%of deformed DNLS equation. We show behavior of analytic
%solutions in numerical simulation. We also show behavior of numerical
%solutions in non-integrable case. 

\section{Integrability of mappings and 
exact localized solutions of the QDNLS equation}
We begin by looking for solutions of the form $\psi_n(t)=\phi_n
e^{-i\omega t}$ where $\phi_n$ is real. Substitution into Eq.(\ref{dnls}),
(\ref{pdnls}) and (\ref{newdnls}) yields the
corresponding symmetric difference equations, respectively
\begin{equation}
 \phi_{n+1}+\phi_{n-1}=(-(\omega/\alpha) +2)\phi_n
-(\beta/\alpha) \phi_n^3,\quad \alpha, \beta \neq 0, \label{rdnls2}
\end{equation}
\begin{equation}
 \phi_{n+1}+\phi_{n-1}=\frac{(-\omega +2\alpha)\phi_n-\beta \phi_n^3}
{\alpha+\gamma \phi_n^2},\quad \alpha,\beta,\gamma \neq 0, \label{rqdnls2}
\end{equation}
and
\begin{equation}
 \phi_{n+1}+\phi_{n-1}=\frac{(-\omega +2\alpha)\phi_n-\beta \phi_n^3}
{\alpha+\gamma \phi_n^2+\delta \phi_n^4},\quad \alpha,\beta,\delta 
\neq 0. \label{rnewdnls2}
\end{equation}
Equation (\ref{rdnls2}) was studied by many authors \cite{RT} and has
chaotic property. 
We apply a singularity confinement (SC) test 
to these difference equations \cite{GRP:SC}. 
Equation (\ref{rqdnls2}) does not pass the SC test except the case in which
Eq.(\ref{rqdnls2}) is reduced to a linear mapping 
$\phi_{n+1}+\phi_{n-1}=(-\beta/\gamma) \phi_n$ 
(in the case of $(\omega-2\alpha)\gamma=\alpha \beta$). 
Thus we may conclude that Eq.(\ref{rqdnls2}) is non-integrable except the case
in which leads to a linear mapping. In the case in which
Eq.(\ref{rqdnls2}) leads to a linear mapping, it has
solutions
$\phi_n
=\frac{1}{\sqrt{C^2-4}}\{
\left(\frac{C+\sqrt{C^2-4}}{C}\right)^n 
\left(\phi_1-\frac{C-\sqrt{C^2-4}}{2}\phi_0\right)
-\left(\frac{C-\sqrt{C^2-4}}{C}\right)^n 
\left(\phi_1-\frac{C+\sqrt{C^2-4}}{2}\phi_0\right)
\}$
where $C=-\beta/\gamma$ and $\phi_0, \phi_1$ are initial values.  

In Eq.(\ref{rnewdnls2}), divergences caused by initial values are
confined to only one-site without losing the initial information for 
certain sets of coefficients. 
That is, under that constraint, Eq.(\ref{rnewdnls2}) has 
the SC property.  
So Eq.(\ref{rnewdnls2}) may be integrable for certain sets of
coefficients. 
Indeed, this equation has soliton type solutions for 
certain sets of parameters.

Substituting $\psi_n=\frac{g_n}{f_n}e^{-i\omega t}$ to
Eq.(\ref{newdnls}), we obtain the multi-linear form
\begin{eqnarray*}
&& (\omega-2\alpha)f_{n+1}f_n^3f_{n-1}g_n
+\alpha f_n^4f_{n-1}g_{n+1}+\alpha f_{n+1}f_n^4g_{n-1}
+\beta f_{n+1}f_nf_{n-1}g_n^3\\
&&\quad +\gamma f_n^2f_{n-1}g_{n+1}g_n^2
+\gamma f_{n+1}f_n^2g_n^2g_{n-1}+\delta f_{n-1}g_{n+1}g_n^4
+\delta f_{n+1}g_n^4g_{n-1}=0. 
\end{eqnarray*}
Using the standard procedure of the Hirota method, 
we obtain the following exact solutions.\\
 \\
{\bf Bright type soliton}\\
\begin{equation}
\psi_n(t)=A\,{\rm sech}(n\log(p)+\log(n_0))e^{-i\omega t},
\end{equation}
where 
$\omega=
\frac{\pm \beta \sqrt{\gamma^2-4\alpha \delta}+4\alpha \delta+\beta \gamma}
{2\delta}$, $p=\frac{\pm \sqrt{\omega (\omega-4\alpha)}+2\alpha-\omega}
{2\alpha}$, 
$A=\pm \frac{1}{2}\sqrt{\frac{\beta \omega (4\alpha-\omega)}
{\alpha \delta (2\alpha-\omega)}}$ and $n_0$ is arbitrary constant.

\begin{figure}
\begin{center}
\includegraphics{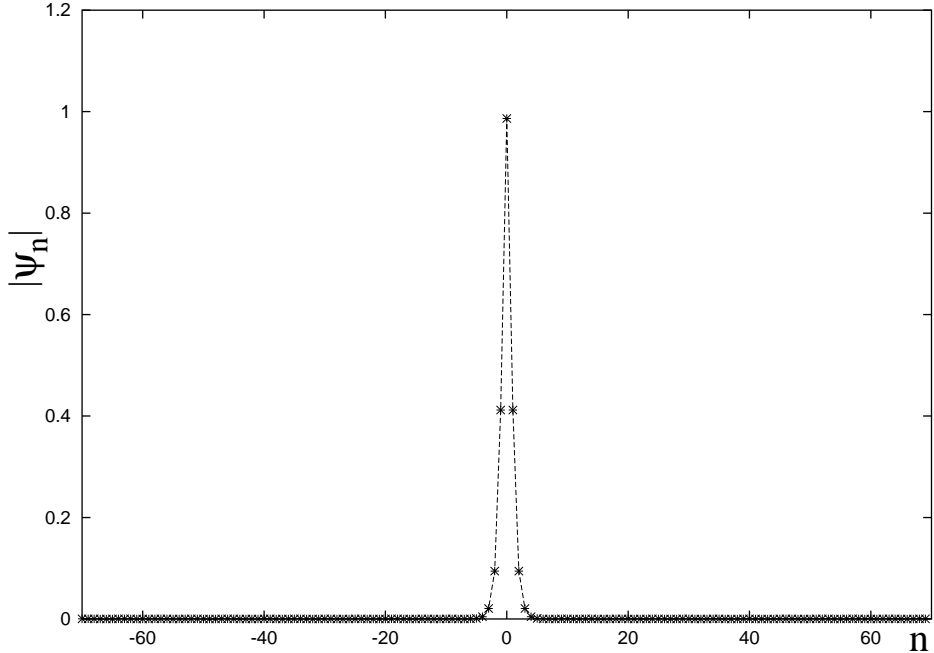}
\caption{Bright soliton of Eq.(\protect \ref{newdnls}). 
$\alpha=1,\beta=0.2,\gamma=4.78701,\delta=-0.2$. 
Time in this snapshot is 100.}
\label{brightdnls}
\end{center}
\end{figure}
\quad\\
{\bf Dark type soliton}
\begin{equation}
\psi_n(t)=A\,{\rm tanh}(n\log(p)+\log(n_0))e^{-i\omega t},
\end{equation}
where 
$\omega
=\frac{\pm \beta \sqrt{\gamma^2-4\alpha \delta}
+4\alpha \delta+\beta \gamma}{2\delta}$,
$p=\sqrt{\frac{\pm \sqrt{2\alpha \omega}+2\alpha+\omega}
{2\alpha-\omega}}$,
$A=\pm \sqrt{\frac{\beta \omega}{2\delta (2\alpha-\omega)}}$ 
and $n_0$ is arbitrary constant. 

The QDNLS equation also has Jacobi elliptic function solutions 
for certain sets of parameters. 

Here, we show Fig.\ref{brightdnls}, which is the result of 
numerical evolution of an initial exact bright soliton.  
We should note that the above localized solution is stable 
for certain parameters. 

\begin{figure}
\begin{center}
\includegraphics{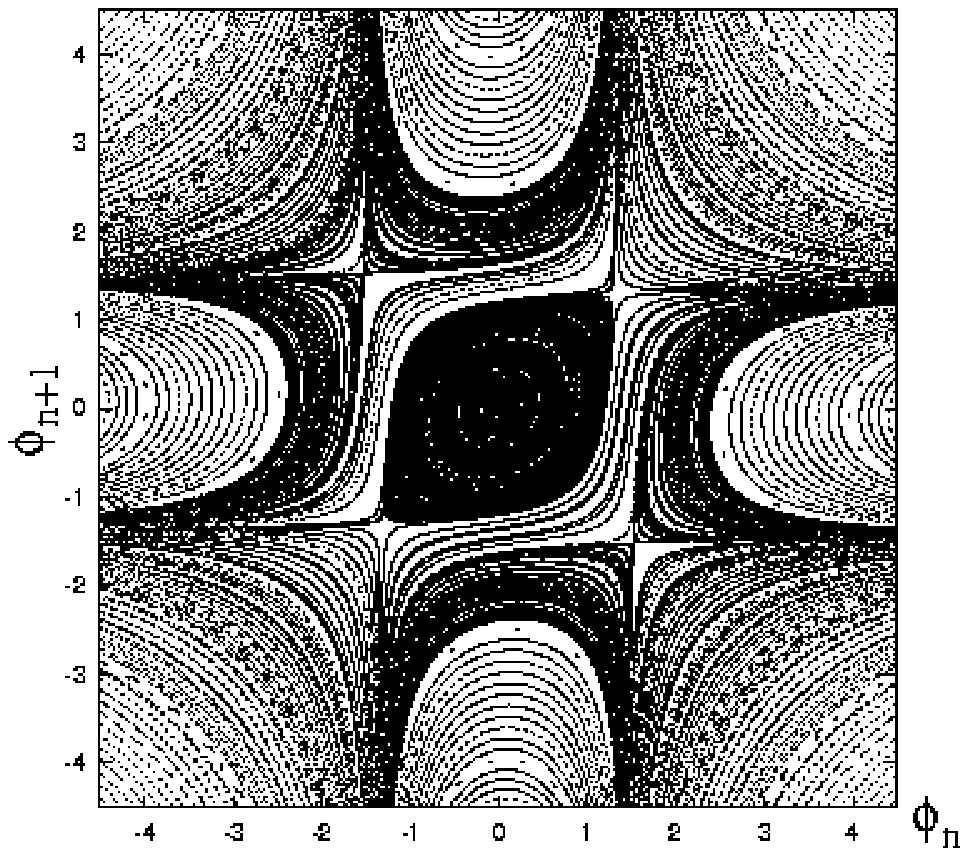}
\end{center}
\caption{A collection of orbits of the map generated by Eq. 
(\protect \ref{rnewdnls2}).(Integrable case:
 $\omega=37, \alpha=22, \beta=7, \gamma=-33, \delta=11$. These parameters
 were determined by the SC test. Each orbits are described by Jacobi
 elliptic functions.)}
\label{order}

\begin{center}
\includegraphics{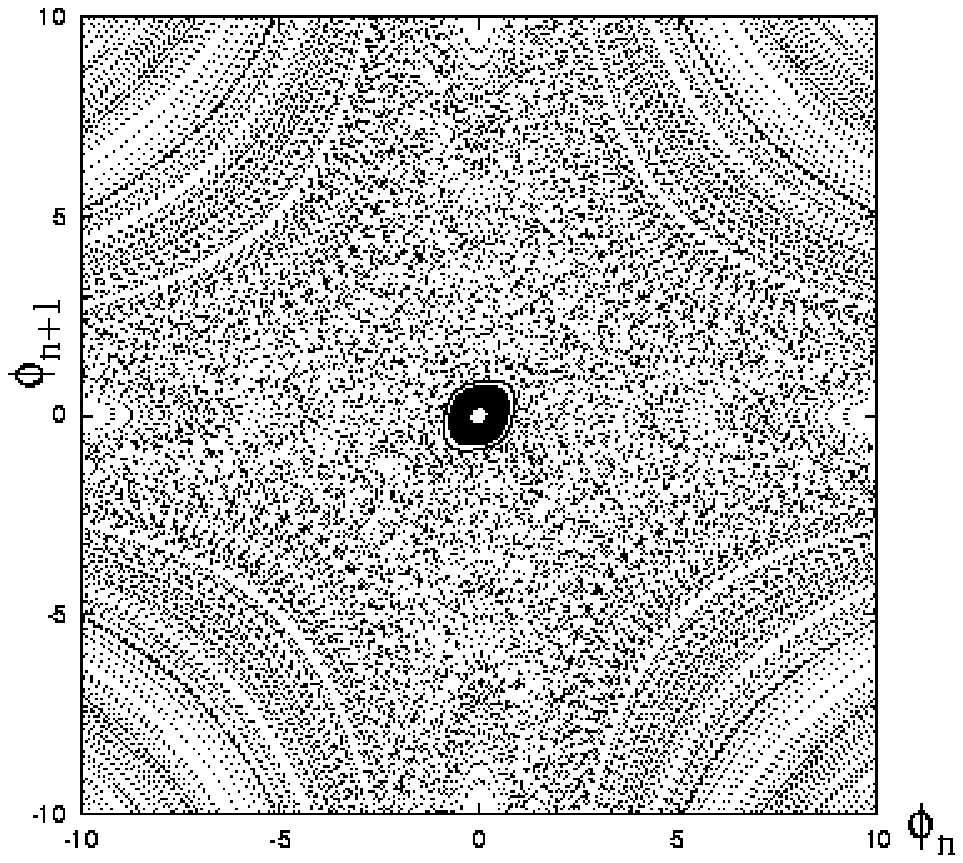}
\end{center}
\caption{A collection of orbits of the map generated by Eq.(\protect
 \ref{rnewdnls2}). (Non-integrable case:
 $\omega=37$, $\alpha=22$, $\beta=6$, $\gamma=-33$, $\delta=11$.)}
\label{qdnlschaos}
\end{figure}

What happens for the sets of parameters for which Eq.(\ref{rnewdnls2}) 
does not pass the SC test? 
We show two figures ( ($\phi_{n+1},\phi_n$) 
phase-plane plot) here. As can be seen, Fig.\ref{qdnlschaos} uses a set of non-integrable parameters and chaotic behavior is obtained 
(compare to Fig.\ref{order}). 

Let us explain why the mapping (\ref{rnewdnls2}) is integrable for
certain sets of parameters. 
It is well-known that the McMillan mapping \cite{McMillan}
$z_{n+1}+z_{n-1}=(A+Bz_n)/(C+Dz_n^2)$ 
has a conserved quantity and could be solved by 
Jacobi elliptic functions \cite{Potts}. 
Equation (\ref{rnewdnls2}) 
for certain sets of parameters reduces to the McMillan
mapping by cancellation of common factors in denominator and numerator
of Eq.(\ref{rnewdnls2}). 
These cases of certain sets of parameters are equivalent to the cases in
which Eq.(\ref{rnewdnls2}) passes the SC test. For example,
in the set of parameters of Fig.\ref{order}, Eq.(\ref{rnewdnls2}) reduces to 
the McMillan mapping
$\phi_{n+1}+\phi_{n-1}=-7\phi_n/(-22+11\phi_n^2)$,
which has an invariant 
$-22(\phi_n^2+\phi_{n-1}^2)+7\phi_n\phi_{n-1}
+11\phi_n^2\phi_{n-1}^2$ 
and Jacobi elliptic function solutions \cite{Potts}. 

We note that the quintic NLS
equation with fourth order dispersion
$iu_t+au_{xx}+\epsilon bu_{xxxx}+c|u|^2u+d|u|^4u=0$
where $\epsilon$ is a small parameter, corresponds to 
a continuous analogue of Eq.(\ref{newdnls}) 
\cite{Dmitriev,Palacios1,Palacios2}.
This equation has exact bright (i.e., $A\,{\rm sech}(kx)e^{it}$ ) and
dark (i.e., $A\,{\rm tanh}(kx)e^{it}$ ) soliton solutions 
for certain sets of parameters \cite{Palacios1,Palacios2,Maruno}. 
The existence of exact localized solutions originates 
from the balance between quintic nonlinear term and 
4th-order derivative term. 
This equation appears in the optical fiber 
transmission studies \cite{Palacios1,Palacios2,AAbook}. 

We can easily generalize to the following DNLS equation:
\begin{eqnarray}
&&i\frac{d\psi_n}{dt}+\alpha (\psi_{n+1}-2\psi_n+\psi_{n-1})
\nonumber\\
&&\quad 
+\gamma_2|\psi_n|^2
(\psi_{n+1}+\psi_{n-1})+\beta_2 |\psi_n|^2 \psi_n\nonumber\\
&&\quad +\gamma_4|\psi_n|^4(\psi_{n+1}+\psi_{n-1})
+\beta_4 |\psi_n|^4 \psi_n\nonumber\\
&&\quad +\cdots\nonumber\\
&&\quad +\gamma_{2N-2}|\psi_n|^{2N-2}(\psi_{n+1}+\psi_{n-1})
+\beta_{2N-2} |\psi_n|^{2N-2} \psi_n\nonumber\\
&&\quad +\gamma_{2N}|\psi_n|^{2N}(\psi_{n+1}+\psi_{n-1})=0.\label{gdnls}
\end{eqnarray}
Using a solution ansatz of the form $\psi_n(t)=\phi_n
e^{-i\omega t}$, where $\phi_n$ is real, 
substitution into equations (\ref{gdnls}) yields the
corresponding symmetric difference equation
\begin{eqnarray}
\phi_{n+1}+\phi_{n-1}
=\frac{(-\omega +2\alpha)\phi_n-\beta_2 \phi_n^3-\beta_4\phi_n^5
-\cdots-\beta_{2N-2}\psi_n^{2N-1}}
{\alpha+\gamma_2 \phi_n^2+\gamma_4 \phi_n^4+\cdots 
+\gamma_{2N-2}\phi_n^{2N-2}+\gamma_{2N}\phi_n^{2N}}.
 \label{rgdnls}
\end{eqnarray}
This mapping has the SC property and low-order degree growth property 
when the coefficients
satisfy some constraints. In fact, it is also possible to construct 
some exact solutions by using standard Hirota method. 
We may consider 
the existence of these exact localized solutions also 
originates from the balance between $2N+1$ order nonlinear term 
and $2N$-order derivative term which comes from 
the higher order term of 
Taylor expansion of second-order difference term. 

\section{Non-integrable mappings and their growth properties}
\begin{figure}
\begin{center}
\includegraphics{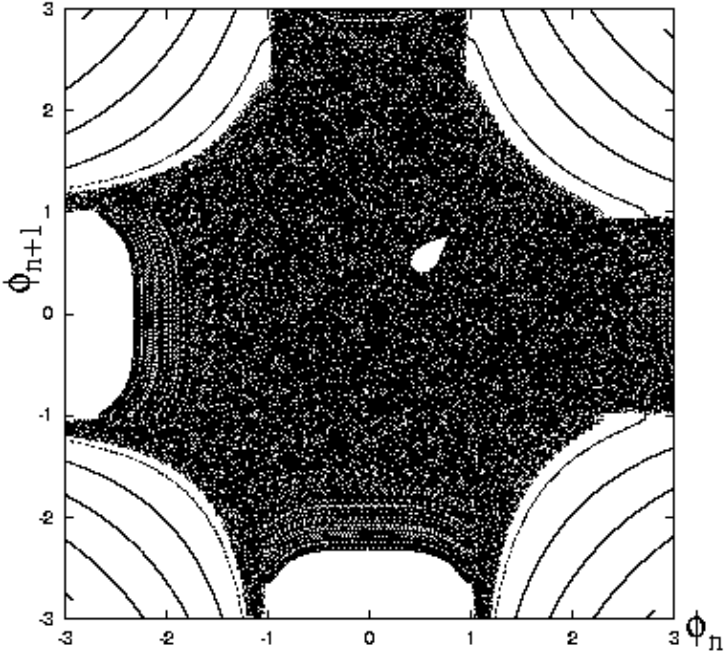}
\end{center}
\caption{A collection of orbits of the map generated by 
Eq.(\protect \ref{dnlsnonint2}).}
\label{chaos}
\end{figure}

Next, we consider the forced quintic discrete 
Nonlinear Schr\"odinger (fQDNLS) equation 
\begin{eqnarray}
&&i\frac{d\psi_n}{dt}+\alpha (\psi_{n+1}-2\psi_n+\psi_{n-1})
+\beta |\psi_n|^2 \psi_n
+\gamma|\psi_n|^2(\psi_{n+1}+\psi_{n-1})\nonumber\\
&&\quad +\delta|\psi_n|^4(\psi_{n+1}+\psi_{n-1})+f\exp(-i\omega t)=0,
\end{eqnarray}
where $f$ is a constant parameter. This equation can be reduced by
$\psi_n=\phi_n\exp(-i\omega t)$ into
\begin{equation}
 \phi_{n+1}+\phi_{n-1}=\frac{f+(-\omega +2\alpha)\phi_n-\beta \phi_n^3}
{\alpha+\gamma \phi_n^2+\delta \phi_n^4}. \label{dnlsnonint}
\end{equation}
Our problem is whether this is integrable or not. 
After we apply the SC test, 
we note Eq.(\ref{dnlsnonint}), for certain sets of 
parameters passes the SC test
without losing initial information. 
For example, we can consider the simple mapping
\begin{equation}
\phi_{n+1}+\phi_{n-1}=1/(1-\phi_n^4).\label{dnlsnonint2}
\end{equation}
This mapping has the SC property.
But this result is a counterexample to the physical intuition that 
nonlinear equations
containing forcing terms are chaotic.  
To verify this, we perform numerical calculation on Eq.(\ref{dnlsnonint2}).
The result obtained from Eq.(\ref{dnlsnonint2}) is shown 
respectively as $(\phi_{n+1},\phi_n)$
phase-plane plots in Fig.\ref{chaos} for varying choices of
$\phi_0,\phi_1$.
(In contrast, $(\phi_{n+1},\phi_n)$
phase-plane plots (Fig.\ref{order}) 
of Eq.(\ref{rnewdnls2}) show periodic orbits.) 

This situation is similar to the case
of Hietarinta and Viallet \cite{HV98}. Algebraic entropy is useful to
check integrability in this case \cite{OTGR,AHH,GTRT}. 
The main argument is that a generic, non-integrable 
mapping has an
exponential degree growth, while integrability is 
associated with low
growth and is typically polynomial. 
The degrees of the iterates of Eq.(\ref{dnlsnonint2}) 
are 0,1,4,17,64,241,900,..., i.e. a
exponential degree growth. Thus we can assert that
Eq.(\ref{dnlsnonint2}) is non-integrable. This fact can be 
confirmed by
considering cancellation mechanism of common factors in 
denominator and
numerator of Eq.(\ref{dnlsnonint2}) (or Eq.(\ref{dnlsnonint})). The
denominator and numerator of Eq.(\ref{dnlsnonint2})
(or Eq.(\ref{dnlsnonint})) do not have common factors. Thus
Eq.(\ref{dnlsnonint2}) (or Eq.(\ref{dnlsnonint})) cannot be reduced to
the McMillan mapping.

From above analysis, we can easily produce the following 
generalized mapping:
\begin{equation}
 \phi_{n+1}+\phi_{n-1}=
\frac{a_0+a_1\phi_n+\cdots+a_{N-1}\phi_n^{N-1}}
{b_0+b_1\phi_n+b_2\phi_n^2+\cdots+b_N\phi_n^N}.\label{gdnlsnonint}
\end{equation}
This mapping also passes the SC test for certain sets of
parameters, but has
exponential degree growth. Thus $(\phi_{n+1},\phi_n)$
phase-plane plots of Eq.(\ref{gdnlsnonint}) also trace chaotic
orbits.

\section{Conclusions}
We have analyzed the QDNLS equation. 
The QDNLS equation is not completely integrable, however 
the QDNLS equation has a number of exact localized solutions 
exists to the QDNLS equation provided that coefficients are bound by
special relations (i.e. partially integrable). 
The set of localized solutions include particular types of solitary
wave solutions, dark soliton solutions and periodic solutions in
terms of Jacobi elliptic functions. 
Clearly these localized solutions represent only a small subset 
of large variety of possible solutions admitted by the QDNLS equation.
Nevertheless, the solutions presented here are found for the
first time and they might serve as seeding solutions for a wider class of
localised structures which exist in this system.
We hope that they will be useful in further perturbative and 
numerical analysis of various solutions to the QDNLS equation.

We mention traveling wave solutions of the QDNLS equation. 
Flach et al. studied moving pulses of the generalized discrete nonlinear
Schr\"odinger equation\cite{Flach2}
\begin{eqnarray}
&&i\frac{d\psi_n}{dt}+\alpha (\psi_{n+1}-2\psi_n+\psi_{n-1})
+F(|\psi_n|^2)(\psi_{n+1}+\psi_{n-1})\nonumber\\
&&\quad \quad \quad +G(|\psi_n|^2) \psi_n=0\,.
\end{eqnarray}
Their result shows that 
the AL system is only generalized discrete 
nonlinear Schr\"odinger equation with sech-type
moving pulses, i.e. 
the QDNLS equation does not have 
exact sech-type exact traveling wave solutions. 
Although the QDNLS equation does not have exact traveling wave solutions, 
there is still the possibility that 
non-exact traveling wave solutions exist. 
Such solutions must conserve the integrals of motion 
(number of soliton, energy and momentum).

We have also found interesting properties of mappings 
derived from the deformed DNLS equation.  

Further studies (e.g. stability of exact localized solutions and 
bound states of multiple pulses, 
and the existence and properties of travelling wave solutions) 
will be addressed elsewhere. 
We believe our result can be applied to the study of discrete
optical (and other) solitons.

KM thanks Y. Kivshar, N. Akhmediev, A. Ankiewicz, K. Kajiwara, S-I. Gotou,
M. Oikawa and T. Tsuchida for useful suggestions. 
KM was supported by a JSPS Fellowship for Young Scientists.
KM and YO thank University of Sydney for warm hospitality. 
NJ's research was supported by the Australian Research Council.

\end{document}